\documentclass[
  journal=pasa,
  manuscript=research-paper, 
  year=2020,
  volume=37,
]{cup-journal}

\usepackage{microtype,siunitx,booktabs}
\usepackage{amsmath}	
\usepackage{amssymb}	
\usepackage{caption}
\usepackage{subcaption}
\usepackage{multicol}
\usepackage{xcolor}

\title{Reduced-resolution beamforming: lowering the computational cost \\ for pulsar and technosignature surveys}

\author{D. C. Price}
\affiliation{International Centre for Radio Astronomy Research, Curtin University, Bentley, WA 6102, Australia}
\alsoaffiliation{SKA Observatory, Science Operations Centre, 26 Dick Perry Avenue, Kensington WA 6151, Australia}
\email[D. C. Price]{dan@seti.au}

\doi{10.1017/pasa.2020.32}

\received {20 Feb 2024}
\revised  {18 Mar 2024}
\accepted {03 Apr 2024}
\published{TBD}


\begin{document}

\begin{abstract}
In radio astronomy, the science output of a telescope is often limited by computational resources. This is especially true for transient and technosignature surveys that need to search high-resolution data across a large parameter space. The tremendous data volumes produced by modern radio array telescopes exacerbate these processing challenges. Here, we introduce a `reduced-resolution' beamforming approach to alleviate downstream processing requirements. Our approach, based on post-correlation beamforming, allows sensitivity to be traded against the number of beams needed to cover a given survey area. Using the MeerKAT and Murchison Widefield Array telescopes as examples, we show that survey speed can be vastly increased, and downstream signal processing requirements vastly decreased, if a moderate sacrifice to sensitivity is allowed. We show the reduced-resolution beamforming technique is intimately related to standard techniques used in synthesis imaging. We suggest that reduced-resolution beamforming should be considered to ease data processing challenges in current and planned searches; further, reduced-resolution beamforming may provide a path to computationally-expensive search strategies previously considered infeasible.

\end{abstract}
\section{Introduction}

Extracting science results from astronomy datasets can often be a computationally-demanding process. Supercomputers have become vital tools in modern astronomy---unfortunately to the point that energy consumption of supercomputers and research infrastructure dominate astronomy's carbon footprint \citep{Stevens:2020,PortegiesZwart:2020}. As well as the ecological impact, access to computing resources is limited by the technology and funding available. Observational astronomers must work within these boundaries to unravel the scientific mysteries and marvels hiding in their data.

In radio astronomy, one of the most computationally-expensive exercises is searching for pulsars and other fast transients. In order to detect a new pulsar with unknown characteristics, one must search across a range of different periods, dedispersion trials, and pulse widths \citep[see Chapter 6,][]{Lorimer:2004}. Searches are particularly challenging when using radio array telescopes, as a large number of beams must be formed to cover a survey area, and the search process must be repeated on each beam. Nevertheless, there are active pulsar searches on several radio array telescopes \citep[e.g.][]{Sanidas:2019,Chen:2021,Singh:2023,Bhat:2023}. Fast Radio Burst (FRB) searches \citep[e.g.][]{Law:2015, Bailes:2017, Bannister:2017, Ng:2017} must also search through multiple dispersion and pulse width trials, but are (somewhat) more manageable as they do not apply a periodicity search.

Another computationally-demanding avenue is the search for extraterrestrial intelligence (SETI), which seeks to detect `technosignatures' from extraterrestrial intelligence as a proxy for intelligent life. Indeed, the SETI@Home project \citep{Anderson:2002} was a pioneer in developing distributed computing in order to attain sufficient compute resources for data analysis. At one stage, the SETI@Home network---peaking at over 5.2\,m volunteers---constituted the largest supercomputer on the planet. 

Decades later, thanks to exponential `Moore's law' scaling of compute capacity, the Breakthrough Listen project \citep{Worden:2017, Isaacson:2017} are able to search petabytes of data for technosignatures with modest server clusters located at the observatory \citep{MacMahon:2018,Price:2018,Lebofsky:2019}, making the distributed computing approach unnecessary. Even so, technosignature searches using radio arrays, such as the Breakthrough Listen program on the MeerKAT telescope \citep{Czech:2021} and the recently-announced  COSMIC\footnote{COSMIC: Commensal Open-Source Multimode Interferometer Cluster Search for Extraterrestrial Intelligence. \url{https://science.nrao.edu/facilities/vla/observing/cosmic-seti}} instrument on the Karl G. Jansky Very Large Array, are orders of magnitude more challenging than previous single-dish searches, motivating new search techniques.

With a radio array, when a desired search strategy is limited by the available compute resources, it is often advantageous to sacrifice sensitivity and imaging fidelity against instantaneous field of view. As beam width scales with $\sim \lambda/D$, where $D$ is the longest baseline in the array and $\lambda$ is wavelength, the number of beams required to cover a given survey area scales with $D^2$. As such, pulsar searches tend to only use the `core' part of an antenna array, and longer baselines are dropped. For example, in the MeerKAT TRAPUM\footnote{TRAPUM: Transients and Pulsars with MeerKAT} survey, only the central 37--41 antennas from the 64-antenna array are regularly used, corresponding to 57--64\% of the peak sensitivity \citep{Chen:2021}. The LOTASS\footnote{LOTASS: LOFAR Tied-Array All-sky Survey} survey used the dense central `Superterp' core of LOFAR to balance field of view against sensitivity. Similarly, the Murchison Widefield Array (MWA) SMART\footnote{SMART: Southern-sky Rapid Two-Metre pulsar survey survey} \citep{Bhat:2023} uses the `compact' MWA configuration of 128 tiles within $\sim$800 m.

The computational challenges for pulsar surveys are exacerbated at low frequencies, as the number of dispersion trials (DMs) required to correct for dispersive delays due to propagation through the interstellar medium scales with $\lambda^2$. The LOTAAS survey searched across 10120 dispersion measure (DM) trials, which required 30M CPU-core hours on the Cartesius supercomputer to process data up to Jan 2019 \citep{Sanidas:2019}. Based on first-pass benchmarks, the SMART survey will require $\sim$60M CPU-core hours to process $\sim$93\,hr of observational data using the OzStar supercomputer \citep{Bhat:2023}. As the SMART team anticipate receiving $\sim$0.5--0.6M CPU-core hours per semester, they are in the process of optimizing and accelerating their search pipeline to ensure  processing is possible.  

In this paper, we introduce a `reduced-resolution' beamforming technique that allows for sensitivity to be traded off against sky coverage. Our reduced-resolution beamforming approach, applied after inter-antenna correlation, down-weights baselines longer than a specified distance. This per-baseline approach results in appreciably larger sensitivity than simply discarding all antennas outside of a core region. Reduced-resolution beamforming effectively covers a desired survey area with a smaller number of beams, for a moderate decrease in survey speed. By doing so, the downstream compute requirements for a given survey can be greatly reduced.

This paper is organized as follows. In Section\,\ref{sec:post-x-bf}, we give an overview of post-correlation beamforming, introduce a compact tensor notation, and highlight the relationship between a power beam and a pixel within a synthesis image. Reduced-resolution beamforming is introduced in Section\,\ref{sec:red-res-bf}, along with survey speed and computational cost metrics. In Section\,\ref{sec:mwa}, we consider the application of reduced-resolution beamforming to the SMART survey; in Section\,\ref{sec:meerkat} we argue that the survey speed for MeerKAT pulsar searches could be increased dramatically by reduced-resolution beamforming, without increasing the number of beams formed. We conclude with a short discussion in Section\,\ref{sec:discussion}.

\section{Post-correlation beamforming}\label{sec:post-x-bf}

\begin{figure*}
    \centering
    \includegraphics[width=0.99\columnwidth]{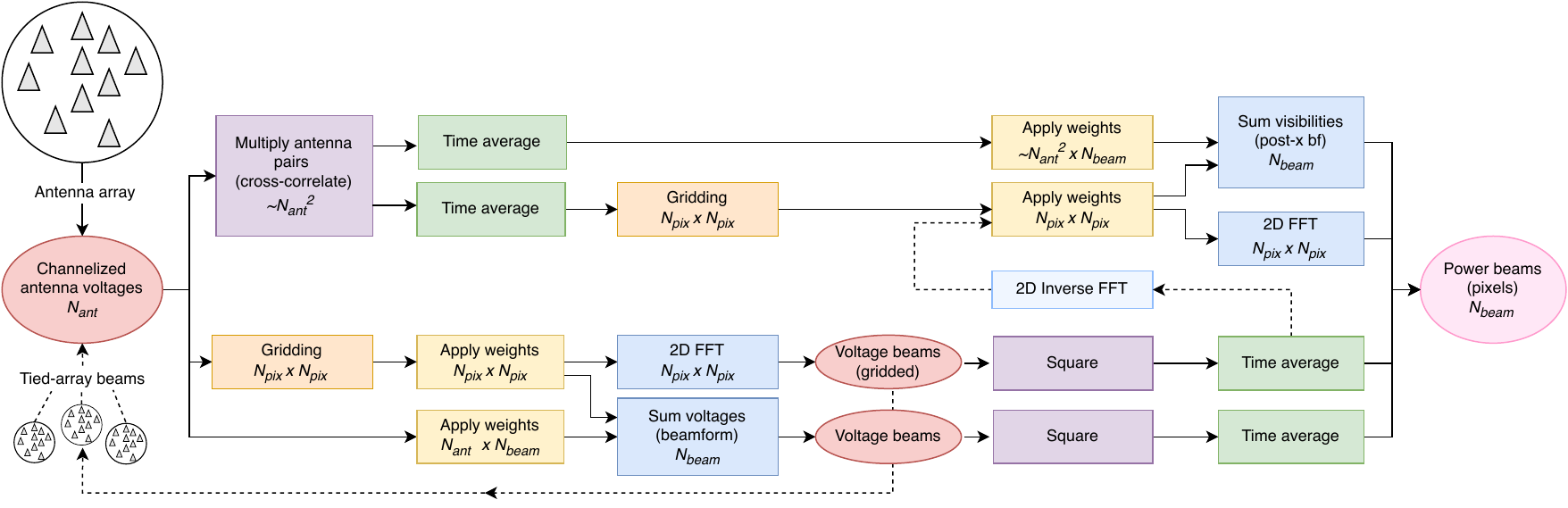}
    \caption{Block diagram showing (simplified) pathways to generate power beams---or equally, images---from antenna voltages (noting $N_{\rm{beam}} \equiv N_{\rm{pix}} \times  N_{\rm{pix}}$). Equivalent tasks are color-coded and italic text corresponds to  data output dimensions; dashed lines represent hybrid architectures.  \textit{\textbf{Standard imaging} }follows the top path (i.e. correlating antenna pairs first): correlations are time-averaged, gridded, weighted, then a 2D FFT is applied to form the image. \textit{\textbf{Post-correlation beamforming} }also follows this top path, but sums visibilities without a gridding step to form a power beam (or multiple beams). \textit{\textbf{Standard tied-array beamforming}} follows the bottom path (i.e. applying weights first): weights are applied to antenna voltages, which are summed to form a voltage beam (or multiple beams). This voltage beam can be squared and time-averaged to create a power beam. \textbf{\textit{Direct imaging correlators}} follow the bottom path too, but apply gridding at the start so a 2D FFT can be used to form a grid of power beams.   }
    \label{fig:pathways}
\end{figure*}

\textit{Tied-array }(coherent) beamforming is a common signal processing operation used to combine the voltage-level outputs from an array of $N$ radio telescopes. A tied-array of $N$ identical telescopes increases sensitivity by a factor of $N$ over a single telescope, but this comes at the expense of a smaller field of view. Multiple tied-array beams can be formed to improve the instantaneous field of view of the array; however, storing and processing multiple beams can be challenging. 

An alternative approach, known as \textit{incoherent beamforming}, sums the output power from each telescope. An incoherent beam retains the field of view, but the sensitivity only improves as $\sqrt{N}$. For computationally-expensive algorithms---such as those required in pulsar, fast radio burst, and technosignature searches---it is often infeasible to form tied-array beams across the instrument's entire field of view. On the other hand, the incoherent beam may not have enough sensitivity nor spatial resolution to be scientifically interesting. 

A beamformed \textit{voltage beam} from $P$ antennas is given by the sum of weights $w(t)=Ae^{i\theta(t)}$
with a voltage stream $v(t)$:
\begin{equation}
    b(t)=\sum_{p=1}^{P}w_{p}(t)v_{p}(t)
\end{equation}
The \textit{power beam}, B, is the beam after squaring and averaging in time. Denoting time average with angled brackets $\langle \rangle$, and treating the weights as static across the time average interval, i.e. $w_p(t)=w_p$,  the power beam is given by

\begin{align}
    B  &= \langle b(t)b^{*}(t) \rangle \ \\
    &= \left\langle\left(\sum_{p=1}^{P}w_{p}v_{p}(t)\right)\left(\sum_{q=1}^{P}w_{q}v_{q}(t)\right)^*\right\rangle \\
    &= \sum_{p=1}^{P}\sum_{q=1}^{P} w_{p} \langle v_{p}(t) v^*_{q}(t) \rangle w^*_{q} \label{eq:basic-postx}
\end{align}

The quantity $\langle v_{p}(t) v^*_{q}(t) \rangle$ can be treated as a  $(P\times P)$ matrix \textbf{V}. This is known as the visibility matrix, and is the fundamental data product produced by a radio interferometer. Equation\,\ref{eq:basic-postx} can alternatively be written in terms of \textbf{V} and a weight vector \textbf{w}:
\begin{equation}
    B = \textbf{w}\,\textbf{V}\,\textbf{w}^H \label{eq:matrix-bf}
\end{equation}
where $\textbf{w}^H$ is the weight vector's Hermitian conjugate. 

In the above, we treated weights as being static; in practice, weights must be updated as a source moves across the sky. The update cadence (i.e. maximum time average interval) required to avoid decorrelation loss depends upon the longest inter-antenna baseline within the array; see \citet{Wijnholds:2018} for further discussion.

There are therefore two paths toward generating a power beam, which we refer to as \textit{pre-correlation} beamforming and \textit{post-correlation} beamforming. These paths are shown diagramatically in Figure\,\ref{fig:pathways}. The figure also shows optional gridding steps, which allows multiple regularly-spaced beams to be formed using a 2-dimensional Fast Fourier Transform (2D FFT) if the input data are regularly-gridded. The dashed lines in Figure\,\ref{fig:pathways} represent hybrid archetictures, e.g. arrays like the MWA and LOFAR where beamforming is performed on a subset of antennas within a `station', and then the station tied-array beams are correlated. 

\subsection{Direct imaging approaches}

With the advent of large-$N$ telescopes, there is increased interest in `direct imaging' instruments which follow the pre-correlation gridded path of Figure\,\ref{fig:pathways}. In one approach, antennas are physically arranged on a grid, then a spatial 2D FFT is applied to form images \citep{Tegmark:2009}. This approach is extended by the Modular Optimal Frequency Fourier (MOFF) imaging technique. In the MOFF approach, antennas do not need to be physically placed on a grid; rather, voltages are gridded electronically based on the aperture illumation of consituent antennas \citep{Morales:2011,Thyagarajan:2017,Kent:2019}. Note that a visibility matrix can be retrieved from a grid of beams via an inverse 2D FFT \citep{Tegmark:2009, Foster:2014}, which is shown in Figure\,\ref{fig:pathways} as a dashed line between pathways.

Here, we focus on non-gridded methods, but note that equivalent tradeoffs between resolution and sensitivity could be leveraged with these alternative architectures.

\subsection{Tensor formalism}

Eq.\,\ref{eq:matrix-bf} can be written in an equivalent compact form using summation notation:
\begin{equation}
    B = \textbf{w}\,\textbf{V}\,\textbf{w}^H  \equiv\  \textbf{w}^{p} \,  \textbf{V}_{pq} \, (\textbf{w}^*)^{q} = \textbf{W}^{pq} \textbf{V}_{pq} \label{eq:post-x-1}
\end{equation}
In this notation, summation is implied over all indices that appear in an upper and lower index (* represents complex conjugation). So, we sum across indexes $p$ and $q$ (summation indices representing antennas) which returns a single value $B$. The quantity  $\textbf{W}_{pq}=\textbf{w}_p \textbf{w}^*_q$ is equivalent the ($P \times P$) matrix formed from the outer product $\textbf{w} \textbf{w}^H$. 

The visibility matrix can itself be written in summation notation, instead of using $\langle \rangle$ brackets, as
\begin{equation}
    \textbf{V}_{p q} = \textbf{v}^{t}_{p} (\textbf{v}^*)_{q t} 
\end{equation}
where $t$ represents time. 

These bold-face quantities are often referred to as \textit{data tensors}. Tensors can simply be considered as multi-dimensional arrays, or generalizations of a matrix into higher dimensions. Here, we will follow the nomenclature used in the TensorFlow software package\footnote{\url{https://www.tensorflow.org/guide/tensor}}. As we are using data tensors as a computational tool, we do not need to worry about other uses of tensors found in physics and mathematics, such as their interpretation as mappings between vector spaces, so we do not define a tensor metric here. We note that tensors have indeed been used previously in calibration and imaging \citep[e.g.][]{Smirnov:2011, Price:2015, Thekkeppattu:2024}.

Whenever a pair of tensors have matching upper and lower indices, entries are summed across all matching indices; this is known as a \textit{pairwise contraction}. The order of pairwise contractions for a chain of tensors is known as the \textit{contraction path}. The computational cost to evaluate a tensor chain depends on the contraction path; finding the optimal path is an NP-hard problem that can quickly become intractable as the number of tensors increases.

\subsection{Computing tensor contractions}

Numerous software packages for performing tensor contractions are available. We highlight that the Python Numpy package \citep{Harris:2020} includes a flexible `one-liner' {\tt einsum} method for computing contractions on Numpy arrays. An equivalent {\tt einsum} method exists in the Cupy software package \citep{cupy:2017}, which offloads computations to a graphics processor unit (GPU). If compiled against the NVIDIA cuTENSOR library\footnote{\url{https://docs.nvidia.com/cuda/cutensor/index.html}}, highly-optimized tensor cores on the GPU may be targeted; these offer orders-of-magnitude greater energy efficiency than regular GPU cores. 

A highly-optimized correlation code that uses tensor cores is detailed in \cite{Romein:2021}, and a complex general matrix multiply (cGEMM) code has been developed for use for beamforming with a phased array feed\footnote{\url{https://gitlab.mpcdf.mpg.de/nesser/tc_cgemm/-/tree/master/}}. Together, these two codes could be used to compute Equation\,\ref{eq:post-x-1}, and the {\tt einsum} method provides a straightforward path for prototyping and developing post-correlation beamforming techniques.

\subsection{Multiple beams and frequency channels}
We can extend Equation.\,\ref{eq:post-x-1} with extra indices (subscripts $l$ and $m$) to represent a grid of $(L\times M)$ beams on the sky, and across $F$ frequency channels (subscript $\nu$):
\begin{equation}
    \textbf{B}_{l m \nu} = \left(\textbf{w}_{l m p \nu} \textbf{v}^{p t}_{\nu} \right)\left( \textbf{w}^*_{l m q \nu} (\textbf{v}^*)^{q}_{\nu t}\right)  \label{eq:beam-grid-pre}
\end{equation}
That is, we sum across indexes $p$ and $q$ (summation indices), and output an N-dimensional array with indices $(l, m, \nu)$. We can also write Equation.\,\ref{eq:post-x-1} as
\begin{equation}
    \textbf{B}_{l m \nu} = \textbf{W}^{p q}_{l m \nu} \, \textbf{V}_{p q \nu}. \label{eq:beam-grid-post}
\end{equation}
If power beams are evaluated at a series of timesteps, we may write $\textbf{B}_{l m \nu}=\textbf{B}(t)_{l m \nu}$ as a function of time.
\begin{equation}
    \textbf{B}(t)_{l m \nu} = \textbf{W}^{p q}_{l m \nu} \, \textbf{V}(t)_{p q \nu}. \label{eq:beam-grid-post}
\end{equation}
There are two key differences between Equation.\,\ref{eq:beam-grid-pre} and Equation.\,\ref{eq:beam-grid-post}. The first is that the tensor $\textbf{W}_{p q l m \nu}$ has $(P^2 \times L \times M \times F)$ entries, but $\textbf{w}_{p l m \nu}$ has only $(P \times L \times M \times F)$ entries. The second is that  Equation\,\ref{eq:beam-grid-pre} requires three pairwise contractions to compute, while Equation\,\ref{eq:beam-grid-post} is only one pairwise contraction---two contractions have already been performed to produce $\textbf{W}_{lmpq\nu}$ and $\textbf{V}_{pq\nu}$. 

To extend to a polarization-aware version, we may simply add a subscript $x$:
\begin{equation}
    \textbf{B}(t)_{l m \nu x} = \textbf{W}^{p q }_{l m \nu x} \, \textbf{V}(t)_{p q \nu x}. \label{eq:beam-grid-post}
\end{equation}
where $x$ represents a set of four polarization coherency measurements (e.g. XX*, XY*, YX*, YY* for a linearly-polarized dual-polarization antenna).

\subsection{Comparison to interferometric imaging}\label{sec:comparison-to-imaging}

In synthesis imaging, an image $I(l,m)$ is created by evaluating 
\begin{equation}
    I(l, m) = \frac{1}{M} \sum_{m=1}^{M} V (u_k, v_k) e^{2\pi (u_k l + v_k m)},
\end{equation}
where $M$ is the number of baselines, and $(u_k,v_k)$ are coordinates relating to the baseline $k$ between a ($p$, $q$) antenna pair \citep[Eq. 7-3,][]{Briggs:1999}. This equation can written as a tensor contraction
\begin{align}
    \textbf{I}_{lm} &= \textbf{W}^k_{lm} \textbf{V}_k \\
    \textbf{W}_{klm} &= \frac{1}{M} exp\left(-i 2 \pi (u_k l + v_k m)\right),
\end{align}
which by splitting $k=pq$ can be rewritten as 
\begin{align}
    \textbf{I}_{lm} &= \textbf{W}^{pq}_{lm} \,\textbf{V}_{pq}, \\
    \textbf{W}_{pqlm} &= \frac{1}{M} exp\left(2 \pi i (u_{pq} l + v_{pq} m)\right).\label{eq:imaging-tensor}
\end{align}
Comparison of Eq.\,\ref{eq:matrix-bf} and Eq.\,\ref{eq:imaging-tensor} reveals that $\textbf{I}_{pq}\equiv \textbf{B}_{pq}$. That is: each pixel in an image is equivalent to a power beam, or alternatively, an image can be created out of a grid of power beams. Note that in synthesis imaging, autocorrelations are generally not included; i.e. the weights tensor $\textbf{W}_{pq}=0$ if $p=q$. From a power beam interpretation, this is equivalent to subtracting the incoherently summed beam from the tied-array power beam \citep{Roy:2018}.

\subsection{Historical perspective}


While the interpretation of image pixels as power beams is not commonly discussed, the link between imaging and beamforming is fundamental to radio astronomy and was leveraged by early interferometers, including the Mills Cross \citep{Mills:1958}. However, for the most part, the historical development of pulsar search pipelines has run parallel to high-fidelity synthesis imaging systems, leading to disparate science communities and techniques.

Since their discovery in 1967 by Jocelyn Bell Burnell, a majority of pulsars have been discovered using single-dish instruments, such as the Parkes Murriyang, Arecibo, Lovell, and Green Bank telescopes \citep{Manchester:2005}. As such, most observing techniques and knowledge within the pulsar community is derived from single-dish approaches. Pulsar search codes such as {\tt PRESTO}\footnote{https://github.com/scottransom/presto} \citep{Ransom:2011} were also designed for single-dish telescopes, and are, in general, incompatible with interferometric data products.

In contrast, radio interferometers have predominantly been used for  long-exposure synthesis imaging. Earth-rotation synthesis, popularized by the Cambridge One-Mile Radio Telescope \citep{Ryle:1962}, combines data across many hours of observation to improve imaging fidelity. Instruments were not designed to output high time resolution data products; rather, they were designed to integrate data for as long as possible to minimize output data rates. Despite their lack of time resolution, pulsars have been detected in synthesis images via other characteristics; exceptionally, the first millisecond pulsar was detected in synthesis images as a source with an anomalous spectrum \citep{Backer:1982}. 

The quest to discover new pulsars has nonetheless led pulsar astronomers toward radio arrays. Modern interferometers have good sensitivity, a large field of view, and better localization capability; on a technical level, tied-array beamforming is increasingly feasible. Fast imaging techniques---where each pixel is treated as a power beam---are have been used for searches of millisecond transient pulses \citep{Law:2015,Law:2018}, as have FFT-based beamforming approaches \citep{Ng:2017}, but are yet to be eagerly adopted in pulsar periodicity searches.

Techosignature searches have followed a similar trajectory. Frank Drake performed the first SETI search using the Tatel single-dish telescope in 1961 \citep{Drake:1961}; single-dish telescopes remained the primary technology used until the construction of the Allen Telescope Array \citep{Welch:2009}. Many technosignature experiments have targeted nearby stars and have sought to maximize frequency coverage \citep[e.g.][]{Isaacson:2017}, hence field of view has not historically been a science driver. But more ambitious surveys of nearby star targets \citep{Czech:2021} motivate faster survey speed, and there is a growing movement toward non-targeted widefield technosignature searches \citep{Houston:2021}.

\section{Reduced-resolution beamforming}\label{sec:red-res-bf}

One tactic used to improve sky coverage of a tied-array beamformer is to only use antennas within a `core' region. Via Equation\,\ref{eq:beam-grid-pre}, selecting a core region is equivalent to setting weights $\textbf{w}_{lmp\nu}$ to zero for any antenna $p$ that is above a cutoff distance from a reference antenna (i.e. \textit{per-antenna} filtering). To first order, the full-width at half-maximum (FWHM)  of the resultant tied-array beam will scale with $\lambda/d_{\rm{max}}$. 

Post-correlation beamforming allows the user to set weights between any antenna pair ($\textbf{W}_{pq}$), allowing \textit{per-baseline} filtering (this is, in fact, the standard approach in synthesis imaging). The per-baseline approach means that all baselines below a given threshold distance can be included in the power beam---including short baselines outside the core. It follows that a per-baseline filtering approach (only possible with post-correlation beamforming) will always yield a more sensitive power beam than a per-antenna approach. 

We refer to post-correlation beamforming schemes that set weights to zero for any baseline above a cutoff distance as \textit{reduced-resolution beamforming}. Controlling the cutoff distance essentially allows for sensitivity to be traded off against beam width. For simplicity, we will only consider weights with magnitude 0 (discard baseline) or 1 (keep baseline); but note there are myriad weighting schemes used in synthesis imaging that could be considered \citep[see Chapter 10,][]{Thompson:2017}. 

\begin{figure}[t]
         \centering
         \includegraphics[width=\textwidth]{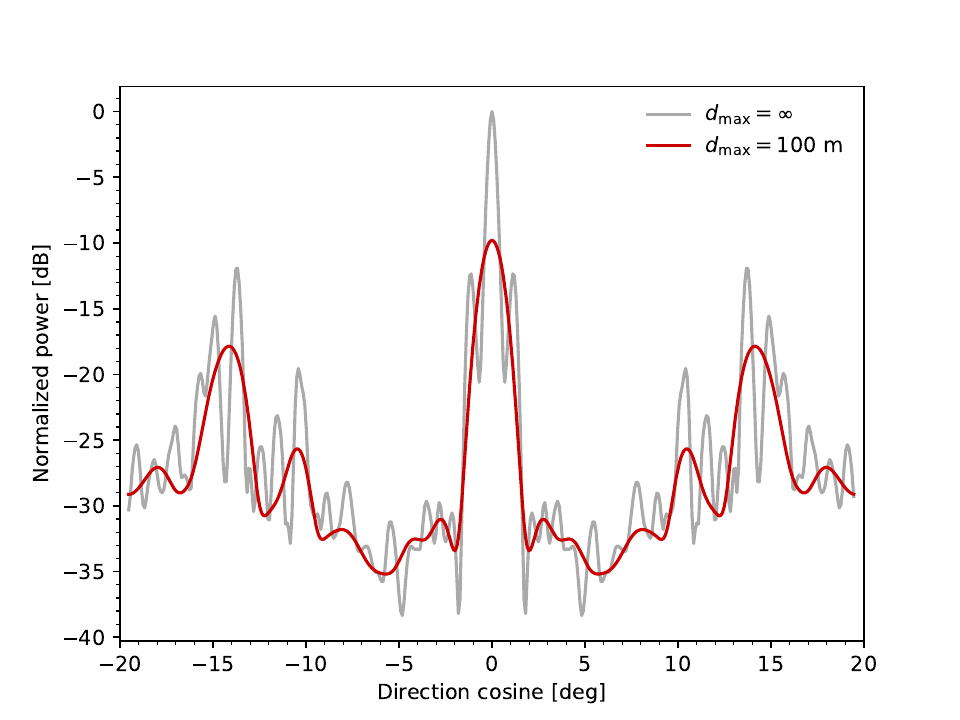}
         \caption{Reduced-resolution beam formed from the MWA compact configuration. The full-resolution beam is shown in grey; a reduced-resolution beam for a 100-m maximum baseline is shown in red. }
         \label{fig:beam-cut}    
\end{figure}

Figure\,\ref{fig:beam-cut} shows an example of reduced-resolution beamforming, as applied to Murchison Widefield Array 128-tile Compact configuration \citep{Wayth:2018}. The grey line shows the beam pattern (known as point spread function, or dirty beam, in synthesis imaging) for a zenith-pointed tied-array beam. The red line shows the resulting beam pattern using reduced-resolution beamforming, with a $d_{\rm{max}}=100$\,m cutoff. The reduced-resolution beam is wider and smoother, but has a lower power gain, which corresponds to lower sensitivity. Note Figure\,\ref{fig:beam-cut} doesn't account for the tile beam pattern.

\subsection{Sensitivity}

From the radiometer equation, the expected thermal noise (in Jy units) of a beam formed via post-correlation beamforming can be written as
\begin{equation}
    \Delta S_{\rm{post-x}} = \frac{k_B T_{\rm{sys}}}{A_e \sqrt{\Delta \nu \tau  N_{\rm{pol}} (2 N_{\rm{baselines}} + N_{\rm{ant}})}} 
    \label{eq:radiometer-postx}
\end{equation}
where we introduce the following quantities:
\begin{center}
     \begin{tabular}{c l}
         $T_{\rm{sys}}$ & System temperature \\
         $k_B$ & Boltzmann's constant \\
         $\Delta\nu$ & Bandwidth \\
         $A_{\rm{ant}}$ & Effective area of an antenna element \\
         $\tau$ & Integration time \\
         $N_{\rm{ant}}$ & Number of antennas \\ 
         $N_{\rm{pol}}$ & Number of polarizations \\
         $N_{\rm{baselines}}$ & Number of baselines in array \\
    \end{tabular}   
\end{center}
Equation\,\ref{eq:radiometer-postx} differs slightly from the standard interferometer radiometer equation \citep[Eq.\,9.26,][]{Wilson:2013} as it includes a $N_{\rm{ant}}$ term to account for autocorrelations. Noting that
\begin{align}
    N_{\rm{baselines}} &= N_{\rm{ant}} (N_{\rm{ant}} - 1) / 2 \\
    N_{\rm{ant}}^2 &= 2N_{\rm{baselines}} + N_{\rm{ant}}
\end{align}
one finds that Equation\,\ref{eq:radiometer-postx} reduces to the tied-array radiometer equation if all baselines are included
\begin{equation}
    \Delta S_{\rm{tied}}  = \frac{2 k_B T_{\rm{sys}}}{A_{\rm{ant}} N_{\rm{ant}} \sqrt{\Delta \nu \tau  N_{\rm{pol}} }}.
    \label{eq:radiometer-tied}
\end{equation}
Additionally, if cross-correlation terms are not included (i.e. $N_{\rm{baselines}} = 0$), then we retrieve the radiometer equation for an incoherent beam
\begin{equation}
    \Delta S_{\rm{incoherent}} = \frac{2 k_B T_{\rm{sys}}}{A_{\rm{ant}} \sqrt{\Delta \nu \tau N_{\rm{ant}} N_{\rm{pol}} }}. \label{eq:radiometer-incoherent}
\end{equation}
If we choose a subset of baselines, based on maximum length, the resulting noise will lie between the ideal $\Delta S_{\rm{tied}}$ thermal noise (all baselines included) and the incoherent $\Delta S_{\rm{incoherent}}$ noise (only autocorrelations included). Put another way, reduced-resolution beamforming provides access to resolution and sensitivity regimes that lie between that of incoherent and coherent beamforming techniques.

\subsection{Fractional sensitivity}

Consider the case where a maximum baseline length $d$ is imposed on an array with longest baseline $D$. For a tied array beamformer, the sensitivity can be treated as a function of baseline length, $\Delta S_{\rm{coherent}}(d)$ (Equation \ref{eq:radiometer-tied}), with the number of antennas in the sub-array is $N'_{\rm{ant}}(d)$. For a post-correlation beamformer (Equation \ref{eq:radiometer-postx}), the thermal noise level reached $\Delta S_{\rm{post-x}}(d)$ depends upon $N'_{\rm{baselines}}(d)$.

It is useful to define a fractional sensitivity factor, $f_{S}$, that relates the sensitivity of the selected subarray to the full array:
\begin{equation}
    \Delta S(d) =  \Delta S(D) \big/ f_{S}.
\end{equation}
For a reduced-resolution beam with all autocorrelations, $N'_{\rm{ant}} = N_{\rm{ant}}$, and
\begin{equation}
    f_{S} = \frac{\Delta S_{\rm{post-x}}(D)}{\Delta S_{\rm{post-x}}(d)} = \frac{\sqrt{2 N'_{\rm{baselines}}(d) + N_{\rm{ant}} } }{N_{\rm{ant}}}
\end{equation}
whereas for a tied-array beam 
\begin{equation}
    f_{S} = \frac{\Delta S_{\rm{tied}}(D)}{\Delta S_{\rm{tied}}(d)} = \frac{N'_{\rm{ant}}(d)}{N_{\rm{ant}}}
\end{equation}
and $N'_{\rm{ant}} \leq N_{\rm{ant}}$. 

\begin{figure}[t]
    \centering
    \includegraphics[width=0.95\columnwidth]{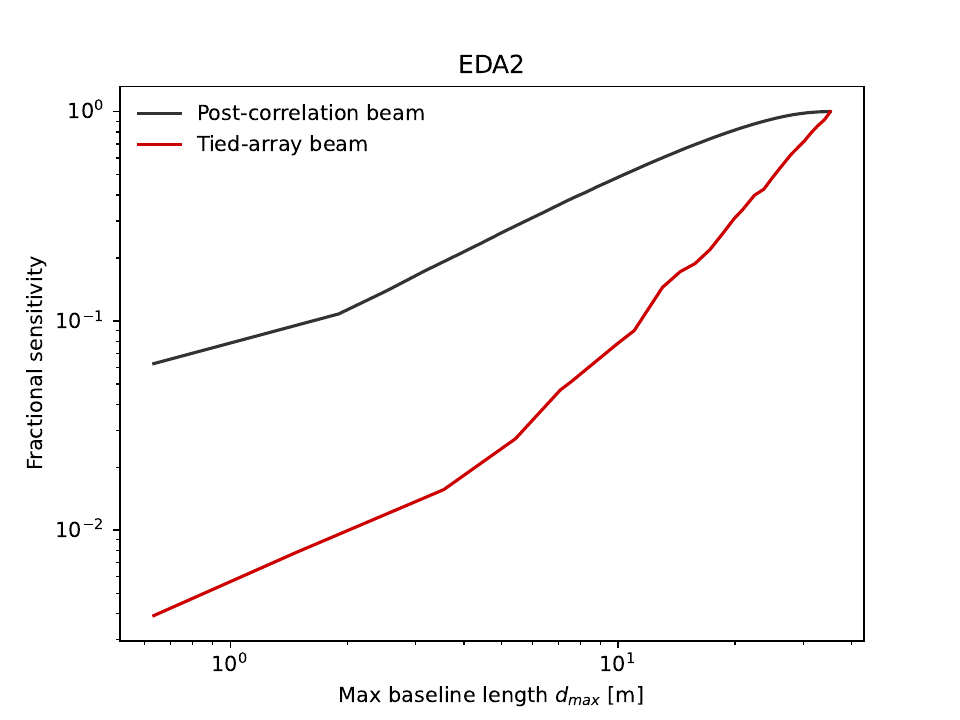}
    \caption{Comparison of EDA2 fractional sensitivity for a tied-array beam (red) and a post-x beamformed beam (black), as a function of maximum baseline length $d_{\rm{max}}$. The post-x approach allows short baselines to be included, boosting sensitivity. }
    \label{fig:tied-vs-postx}
\end{figure}

Figure \ref{fig:tied-vs-postx} shows the fractional sensitivity for a reduced-resolution beam and tied-array beam as a function of maximum baseline length, using the EDA2\footnote{EDA2: Engineering Development Array 2} array as an example \citep{Wayth:2021}.  The reduced-resolution beam always has a larger fractional sensitivity than a tied-array beam from a sub-selection of antennas. This fact is a key motivation for reduced-resolution beamforming.

\begin{figure*}
     \centering
     \begin{subfigure}[b]{0.31\textwidth}
         \centering
         \includegraphics[width=\textwidth]{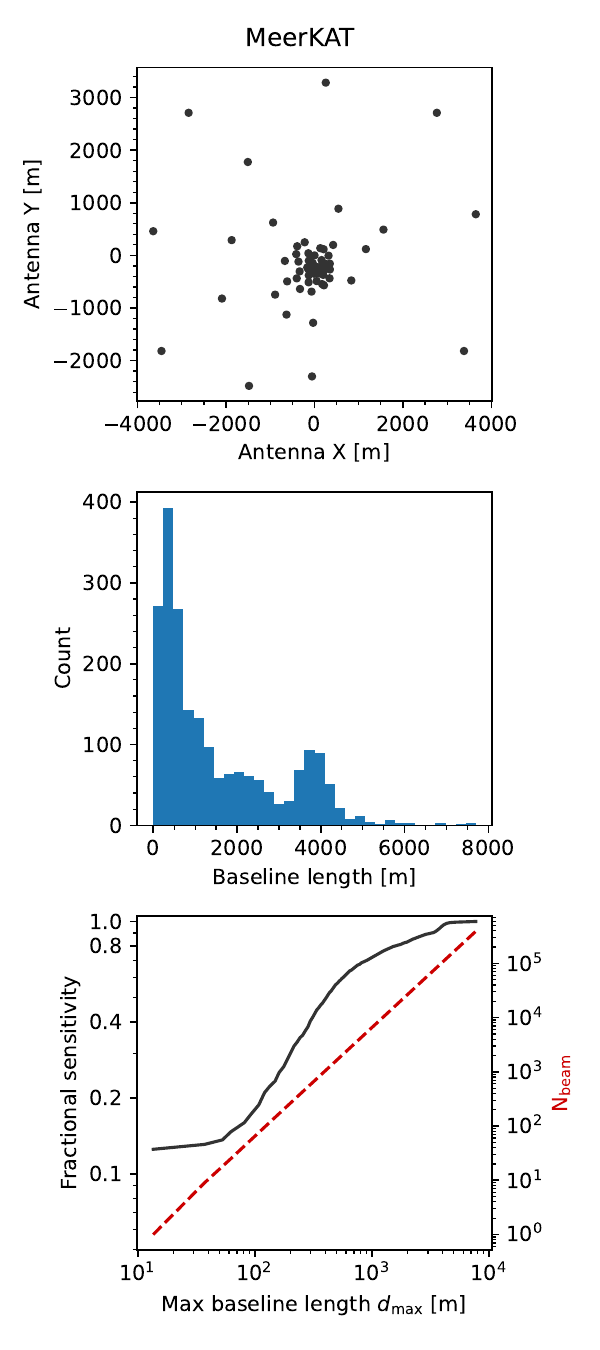}
         \caption{MeerKAT}
         \label{fig:y equals x}
     \end{subfigure}
     \hfill
     \begin{subfigure}[b]{0.31\textwidth}
         \centering
         \includegraphics[width=\textwidth]{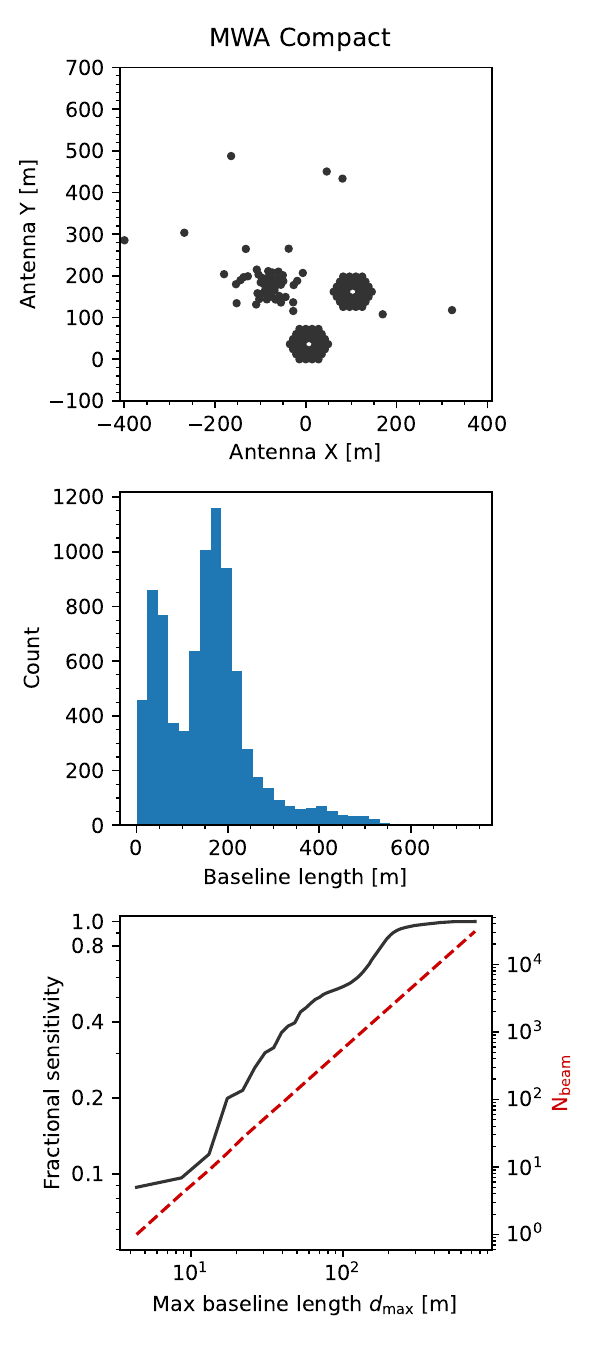}
         \caption{MWA (compact configuration)}
         \label{fig:three sin x}
     \end{subfigure}
     \hfill
     \begin{subfigure}[b]{0.31\textwidth}
         \centering
         \includegraphics[width=\textwidth]{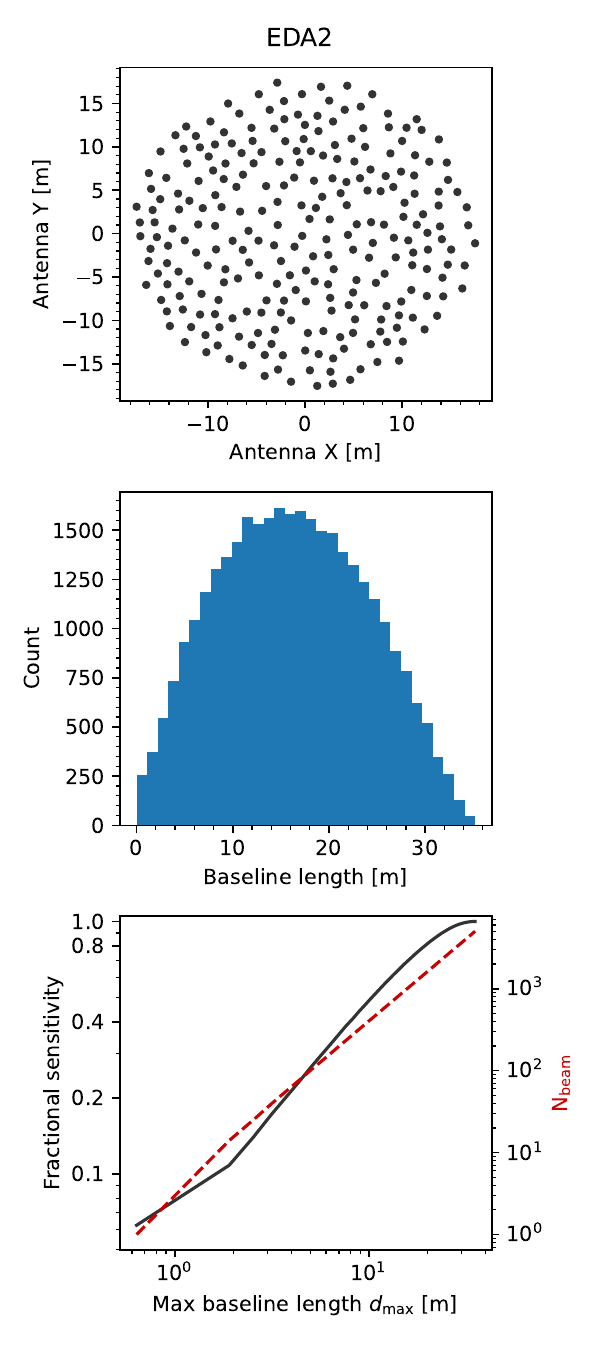}
         \caption{EDA2 (SKA-low pathfinder)}
         \label{fig:five over x}
     \end{subfigure}
        \caption{Sensitivity analysis for reduced resolution beamforming applied to MeerKAT, the MWA (compact configuration) and the EDA2. The top panels show antenna location, and the middle panels show histograms of baseline lengths for each telescope. The bottom panel shows the fractional sensitivity for reduced resolution beamforming, as a function of maximum baseline length (black lines). The minimum sensitivity is equivalent to incoherent beamforming, and maximum sensitivity is equivalent to standard beamforming. The number of beams required to cover each telescope's field of view scales with $d_{\rm{max}}^2$ (dashed red lines).}
        \label{fig:sensitivity-analysis}
\end{figure*}

The slope of $\Delta S_{\rm{post-x}}(d)$ depends upon the antenna configuration. Figure \ref{fig:sensitivity-analysis} considers the application of reduced-resolution beamforming to three radio arrays: the 64-antenna MeerKAT telescope \citep{Camilo:2018}, the 128-tile MWA  \citep[in compact configuration,][]{Wayth:2018}, and the EDA2 \citep{Wayth:2021}. The top panels show the antenna layout for each array, and the middle panels show baseline distribution histograms as a function of baseline length. The lower panel shows the fractional sensitivity for each array as a function of maximum baseline length (solid black line), and the number of beams required to fill the primary field of view, assuming the beam width scales as $\lambda/d$ (dashed red line).

A key takeaway from Figure \ref{fig:sensitivity-analysis} is that all telescopes reach a reasonable fractional sensitivity ($\sim0.5$) with an order of magnitude fewer beams to fill their field of view than the full array. In the coming sections, we will explore how this finding could be applied on these telescopes. 

\subsection{Survey speed}

Survey speed is a useful metric for optimizing the performance of an array. The point-source survey speed figure of merit (PFoM) relates to the time taken to survey a field with solid angle $\Omega_{\rm{survey}}$ to a required thermal noise level in Jy, $\Delta S_{\rm{survey}}$. Following Equation 3 of Cordes (2009)\footnote{\url{https://www.skatelescope.org/uploaded/26865_109_Memo_Cordes.pdf}}, we may define a PFoM modified in a similar fashion to \citet{Chen:2021} to account for number of beams and maximum baseline length. 

The fraction of the maximum PFoM (all antennas, beams covering instrumental FoV) reached with a reduced-resolution beamforming system is
\begin{equation}
    f_{\rm{PFoM}} = \frac{{\rm{PFoM}}(d)}{\rm{PFoM_{full}}} = f^2_S \times f_{\rm{FoV}}
\end{equation}
where the fractional coverage of the instrument's field of view ($\Omega_{\rm{full}}$), depends on the number of beams formed:
\begin{equation}
    f_{\rm{FoV}} = \frac{\Omega_{\rm{FoV}}(d)}{\Omega_{\rm{full}}}  \approx \frac{N_{\rm{beam}}}{\Omega_{\rm{full}}} \left(\frac{\lambda}{d} \right)^2,
\end{equation}
assuming resolution follows the $\lambda/d$ rule of thumb. However, the actual beam width depends upon both the array configuration and the beam's declination in a complex fashion. Many radio arrays have decreasing antenna density for longer baselines, for which $\lambda/d$ underestimates the beam width. Nevertheless, if the number of beams is fixed, we expect the fractional field of view to decrease rapidly as $d$ increases.

\subsection{Computational cost}

 The approximate computational cost of a correlator and multi-pixel beamformer can be written in terms of complex-multiply accumulate operations per second (CMACs/s) as
\begin{align}
   {\rm{C_X}} &=  \left(N_{\rm{ant}} N_{\rm{pol}}(N_{\rm{ant}} N_{\rm{pol}} + 1)\right) \times \Delta \nu  \\ \label{eq:cmac-x}
   {\rm{C_B}} &=  \left(N_{\rm{beam}} \, N_{\rm{ant}} \,  N_{\rm{pol}} \right) \times \Delta \nu
\end{align}
where $\Delta\nu$ is the bandwidth of data processed.
To form post-correlation beams from the correlator output requires an additional step, with computational cost
\begin{equation}
   {\rm{C_{pxb}}} = f_{\rm{baselines}} \times \left(  N_{\rm{pxb}} N^2_{\rm{ant}} N^2_{\rm{pol}} \right) \times \frac{\Delta \nu}{N_{\rm{int}}} \label{eq:cmac-pxb}
\end{equation}
where $N_{\rm{int}}$ is the number of time samples summed during integration,  $f_{\rm{baselines}}$ is the fraction of baselines included, and $N_{\rm{pxb}}$ is the number of post-correlation beams. Note that the cost of FFT-based imaging scales proportionally to $N_{\rm{pxb}} {\rm{log}}_2 N_{\rm{pxb}}$, and for most arrays\footnote{Noting $N_{\rm{pxb}} \equiv N_{\rm{pix}} \times N_{\rm{pix}}$, where $N_{\rm{pix}}$ is image width in pixels.}
\begin{equation}
    {\rm{log}}_2 N_{\rm{pxb}} \ll N_{\rm{ant}}^2,
\end{equation}
so we stress that the PXB approach is generally far more computationally expensive than the 2D-FFT approach employed in imaging pipelines. That said, for fair comparison with computationally expensive searches---where only a few beams can be processed on any one compute server---we do not compare against FFT-based imaging any further in this article. 

Based on Equations \ref{eq:cmac-x}--\ref{eq:cmac-pxb}, tied-array beamforming is more computationally expensive if $N_{\rm{beams}} \lessapprox N_{\rm{pxb}}$ and/or if the correlator does not integrate many time samples, i.e. $N_{\rm{int}}$  is large.

This conclusion is consistent with \citet{Roy:2018}, which considers post-correlation beamforming techniques for pulsar studies with the Giant Metrewave Radio Telescope. \citet{Roy:2018} also concludes that post-correlation beamforming is computationally cheaper for a large number of beams at low time resolution, and regular beamforming is cheaper for a small number of high time-resolution beams. 

As with imaging, the cost of post-correlation beamforming (Eq \ref{eq:cmac-pxb}) can be reduced significantly if visibilities are gridded and a 2D FFT is used \citep{Briggs:1999}. However, the FFT approach is not be suitable for a small number of beams and/or wide-field imaging; further, when distributing processing tasks across a supercomputer, it may be advantageous to have only a small number of beams per node. As such, we present values for the non-gridded approach.

\begin{table}[t]
\caption{MWA SMART survey parameters for a single zenith pointing.\label{tab:smart-params}}
\begin{center}
     \begin{tabular}{c c l}
     \hline
         $N_{\rm{ant}}$ & 128 &  Number of antennas \\ 
         $N_{\rm{pol}}$ & 2 & Number of polarizations \\
         $N_{\rm{chan}}$ & 3072 & Number of channels \\
         $\Omega_{\rm{FoV}}$ & 610 & Field of view \\
         $N_{\rm{beam}}$ & 6100 & Number of beams to tile FoV \\
         $\Delta \nu$ & 30.72  & Bandwidth (MHz) \\
         $\tau$ & 100  & Integration time ($\mu$s) \\
         $t_{\rm{obs}}$ & 80 & Observation time (min) \\
    \hline
    \end{tabular}   
\end{center}
\end{table}
\vspace{0.5 cm} 

\section{Application to MWA SMART}\label{sec:mwa}

SMART \citep{Bhat:2023} is a pulsar and fast transient search project on the MWA, using the 128-tile compact configuration \citep{Wayth:2018}. SMART aims to conduct a pulsar search across the full sky below 30$^\circ$ declination, to a limiting sensitivity (10$\sigma$) of 2--3 mJy. Observations began in 2018, and are approximate 75\% complete \citep{Bhat:2023b}. The SMART project stores voltage-level data products for each of 128 tiles within the MWA compact configuration, which allows beamforming and/or correlation to be performed offline. SMART consists of 70 pointings of 80-minute duration (93 hr total), which corresponds to $\sim3$\,PB of data. These data are stored on the Pawsey Banksia object store\footnote{\url{https://pawsey.org.au/systems/banksia/}}.

SMART uses a multi-pixel tied-array beamformer \citep{Swainston:2022}, which is used to form up to 20 beams at once (per compute node). Each pointing requires $\sim6000$ beams to cover the full field of view. The pulsar search is performed on power beam data (Stokes-I). On the OzSTAR supercomputer\footnote{\url{https://supercomputing.swin.edu.au/}}, beamforming 10 minutes of data takes 2 kSU (service units), and postprocessing search tasks take 25 kSU (1 kSU is equivalent to 1000 CPU-core hours, or 125 GPU-core hours). Extrapolating from these numbers, assuming processing scales linearly with observation time, processing all 70 pointings for the full 80 minutes would require $\sim1120$ kSU for beamforming, and $\sim14,000$ kSU for post-processing. 

\subsection{Trading sensitivity against compute requirements }
The post-processing requirements for SMART scale proportionally with the number of beams searched. By applying reduced-resolution beamforming, the SMART survey volume can be searched with fewer beams, potentially improving computational requirements by an order of magnitude for a moderate drop in sensitivity. However, any improvement in downstream processing costs must be offset against any increases in beamforming costs. Using Equations \ref{eq:cmac-x}--\ref{eq:cmac-pxb}, and the SMART survey parameters (Table \ref{tab:smart-params}), the computational costs are:
\begin{align}
    {\rm{C_X}} &= 16.2 \ {\rm{TCMAC/s}} \\
    {\rm{C_B}} &= 767.6 \ {\rm{TCMAC/s}}.
\end{align}
The comparable computational cost for reduced-resolution beamforming ($C_{\rm{PXB}}$) as a function of fractional sensitivity $f_s$ is shown in Table \ref{eq:post-x-stats}. The computational cost is considerable, as the integration time $\tau=100 \mu$s only allows $N_{\rm{int}}=2$. For such a large number of beams, computational cost could be reduced significantly by using a 2D FFT imaging approach. Regardless, Selecting $f_s=0.5$ yields $C_{\rm{PXB}}=1146$\, TMAC/s, i.e. roughly 1.5$\times$ the cost of tied-array beamforming, and the current SMART tiling approach could be maintained.

\begin{table}[t]
\caption{Computational requirements and corresponding sensitivity limits for MWA reduced-resolution beamforming, for increasing fractional sensitivity. \label{eq:post-x-stats}}
\begin{center}
     \begin{tabular}{c c c c c c}
     \hline
     $f_s$ & $d_{\rm{max}}$ & $N_{\rm{beam}}$ & $N_{\rm{baseline}}$ & $C_{\rm{PXB}} $  & $\Delta S $ (10$\sigma$)\\
           &   (m)          &                 &                     &  (TCMAC/s)       &  (mJy) \\ 
    \hline
	0.2 & 17.4 & 	16 & 	 389 & 12    & 10.0--15.0\\
	0.3 & 30.8 & 	53 & 	 810 & 84    & 6.7--10.0\\
 	0.4 & 48.3 & 	130 & 	1353 & 343   & 5.0--7.5\\
 	0.5 & 70.4 & 	277 & 	2121 & 1146  & 4.0--6.0\\
 	0.6 & 123.2 & 	851 & 	2970 & 4931  & 3.3--5.0\\
 	0.7 & 154.0 & 	1330 & 	4084 & 10596 & 2.9--4.3\\
 	0.8 & 180.3 & 	1825 & 	5368 & 19112 & 2.5--3.8\\
 	0.9 & 211.1 & 	2501 & 	6643 & 32412 & 2.2--3.3\\
    \hline
    \end{tabular}   
\end{center}
\end{table}

While the beamforming cost is higher, the number of beams to search drops from 6100 to 277. It follows that post-processing requirements will drop by a factor of 22$\times$, while sensitivity drops only by a factor of 2$\times$. We extrapolate that the post-processing requirements on OzSTAR would drop from $\sim14,000$ kSU to $\sim636$ kSU. The SMART team anticipates securing 500--600 kSU per annum \citep{Bhat:2023}, meaning it is plausible that the entire survey could be processed within a year if reduced-resolution beamforming is adopted (instead of an implausible 22 years).

The processing requirements for SMART may also be alleviated by planned improvements to the performance of their pulsar search pipeline. The SMART pipeline currently uses {\tt PRESTO}, a CPU-only code written primarily in ANSI C \citep{Ransom:2011}. GPU search codes, such as {\tt astroacclerate} \citep{Armour:2020} and {\tt peasoup} \citep{Barr:2020}, are hoped to offer an order-of-magnitude improvement. For example, the {\tt astroacclerate} Fourier domain acceleration search is up to $8\times$ faster than its {\tt PRESTO} equivalent \citep{Adamek:2020}; recent work to exploit half-precision data types may offer an additional $1.6\times$ speedup \citep{White:2023}. Alleviating I/O bottlenecks, such as disk read speed, may also improve efficiency. Regardless, any speedup from pipeline optimization is complementary to the reduced-resolution beamforming approach.

\subsection{Application to technosignature searches}

A narrowband technosignature search of SMART data has been proposed, which would be the first all-sky technosignature search in the Southern Hemisphere. As such, a SMART technosignature search would place some of the most stringent constraints on the prevalence of putative engineered transmitters in the Galaxy. It is argued that all-sky SETI searches at low frequency are one of the most compelling methods to detect evidence of technologically-capable life beyond Earth \citep{Garrett:2017,Houston:2021}. Narrowband technosignature searches do not require high time resolution, which decreases the computational cost of post-correlation beamforming. For example, decreasing the time resolution from $100\,\mu$s to 1\,s decreases $C_{\rm{PXB}}$ (Equation\,\ref{eq:cmac-pxb}) by a factor of 10,000. 

However, narrowband technosignature searches require high frequency resolution, meaning that large data volumes must be stored in memory on which a large FFT can be performed. This will limit the number of beams that can be processed at any time. As with the pulsar search, the post-processing requirements for narrowband technosignatures scale linearly with the number of beams. Similar tradeoffs between sensitivity and number of beams are thus well motivated for technosignature surveys.

\section{Application to MeerKAT: optimizing survey speed}\label{sec:meerkat}

The TRAPUM program \citep{Chen:2021} is a tied-array pulsar search program on the MeerKAT 64-antenna array \citep{Camilo:2018}. Pulsar searches are conducted using up to $\sim 1000$ power beams, formed from a tied-array beamforming approach \citep{Barr:2018}. An incoherent beam may also be formed, and the maximum number of beams can increase to 4096 if a smaller subset of antennas is beamformed. 

\begin{figure}[t]
    \centering
    \includegraphics[width=0.95\columnwidth]{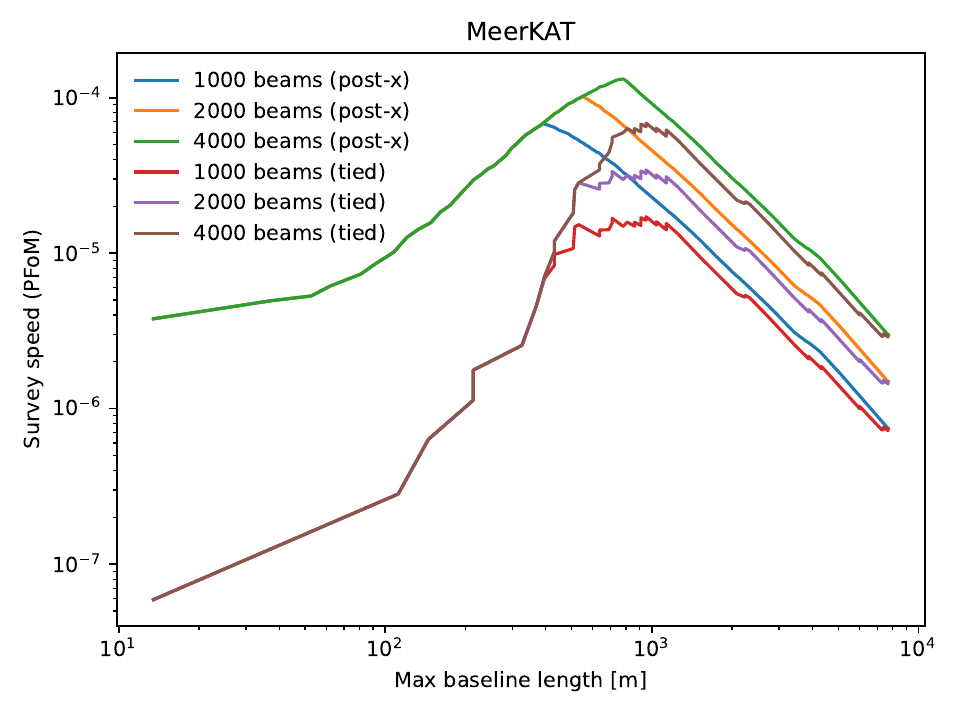}
    \caption{ Comparison of MeerKAT survey speed with reduced resolution beamforming and tied-array beamforming approaches. }
    \label{fig:meerkat-tied-vs-postx}
\end{figure}

TRAPUM uses a point-source figure of merit (PFoM) modified for MeerKAT to optimize survey speed \citep[Equation 4][]{Chen:2021}, from which the optimal number of antennas is found to be between 37 and 41, depending on the altitude observed; this corresponds to a $\sim1$\,km maximum baseline, and 58--64\% fractional sensitivity. If reduced-resolution beamforming was applied, the corresponding fractional sensitivity would be 71--75\% for the same baseline cutoff. The extra baselines would also slightly improve the beam sidelobe response by filling in the $uv$-plane. 

Based on $N_{\rm{ant}} = 64$, and time integration used in the TRAPUM search mode ($\tau = 76\mu$s, $N_{\rm{int}} = 32$), the computational cost of reduced-resolution beamforming yields ${\rm{C_{PXB} = 4 C_{B}}}$. Memory requirements may also be higher, which could limit the number of reduced-resolution beams that could be formed. 

Figure\,\ref{fig:meerkat-tied-vs-postx} shows a comparison of survey speed against maximum baseline length for the MeerKAT array. If tied-array beamforming is used, the maximum baseline length corresponds to the size of the `core' region of antennas used. For reduced-resolution beamforming, the maximum baseline length corresponds to the subset of baselines from the entire array. Note the PFoM for the tied-array beam is comparable to Figure 6 \citet{Chen:2021}. The PFoM is consistently higher for reduced-resolution beamforming. For $N_{\rm{beam}}=1024$, the maximum PFoM peaks 3.9$\times$ higher for the reduced-resolution approach. 

However, taking into account the instrument's capabilities, we conclude that post-correlation beamforming is unlikely to yield a higher PFoM. TRAPUM processes data streams in real time, and the beamformer is computationally-bound. So, the increase in computational cost (${\rm{C_{PXB} = 4 C_{B}}}$) outweighs the benefit of the 3.9$\times$  PFoM increase: 4$\times$ fewer beams could be formed. Nevertheless, the approach may be useful for other projects on MeerKAT where the downstream search algorithms are computationally expensive.

\section{Conclusions}\label{sec:discussion}

This article has introduced reduced-resolution beamforming: a post-correlation beamforming approach that allows sensitivity to be traded against the number of beams needed to cover a survey area. Reduced-resolution beamforming offers an alternative to tied-array beamforming of a core region of antennas. The sensitivity of a reduced-resolution beam is always higher than tied-array beamforming of a core antenna region.

There are two pathways toward forming power beams: pre-correlation beamforming, in which voltages are summed, squared then averaged, and post-correlation beamforming, in which antenna pairs are cross-correlated before summation and averaging (Figure\,\ref{fig:pathways}). Our approach relies on the latter, in which per-baseline weights can be applied in lieu of per-antenna weights. 

In Section 2 we introduced a tensor formalism to highlight the difference between pathways of Figure\,\ref{fig:pathways}, and showed that a pixel within an interferometric image is equivalent to a power beam. Tensors have become standard data structures within commonly-used software packages such as Numpy, and offer compact notation; given these advantages, we encourage the use of tensor formalisms for further research on beamforming approaches. 

There are two main advantages of reduced-resolution beamforming. Firstly, by decreasing the number of beams needed to cover a survey area, the downstream  processing requirements to search the beams is lower: in Section\,\ref{sec:mwa} we show by applying reduced-resolution beamforming to the SMART survey, the same sky volume could be covered with 22$\times$ fewer beams, while retaining 50\% fractional sensitivity. Secondly, for a fixed number of beams, the survey speed of the telescope may be increased: in Section\,\ref{sec:meerkat} we argue that the PFoM for the TRAPUM search mode on the MeerKAT telescope could be increased by 3.9$\times$ by if reduced-resolution beamforming was adopted, albeit at the expense of increased computational requirements for beamforming.


While reduced-resolution beamforming demands a sensitivity tradeoff, it could be offset by improved downstream processing. For example, in a pulsar search, more dispersion measure or acceleration trials may become feasible if fewer beams need to be searched, improving the final signal-to-noise of candidates. One could also consider using reduced-resolution beamforming as a first-pass approach to find candidates above a less stringent threshold (e.g. 5$\sigma$), then form tied-array beams to follow-up candidates, setting a more stringent threshold (e.g. 10$\sigma$) for validation or rejection.     

Nevertheless, there are limitations of the approach. Reduced-resolution beamforming produces power beams, so cannot be applied in cases where voltage beams are needed (e.g. coherent dedispersion). For science cases where confusion noise is a relevant concern, such as continuum imaging, reduced-resolution beamforming is not appropriate. Another disadvantage is that the decrease in angular resolution means that any transient object found will have a comparatively poor localization, although this could be improved using intra-beam methods \citep{Obrocka:2015}. Finer localization of candidate objects could also be done by reprocessing/reobserving at full resolution, then using known parameters (e.g. DM, period, pulse width) to narrow the search space. Finally, the computational requirements for post-correlation beamforming may become unfeasibly large for arrays with a large number of antennas (Equation\,\ref{eq:cmac-pxb}). Memory requirements for storing the correlation matrix and beamforming weights may also  limit the number of beams that a single compute node could process. 

All things considered, we conclude that reduced-resolution beamforming is best suited to science cases where downstream processing requirements dominate the processing budget, such as searches for pulsar, technosignatures and fast transients. We encourage experimental application and verification of the approach to ease the data processing challenges faced by current and planned searches.



\begin{acknowledgement}
D.C. Price thanks E. Barr, R. Wayth, N. Thyagarajan, C. Bassa and the SMART collaboration for their comments.
\end{acknowledgement}

\paragraph{Software} Numpy \citep{Harris:2020}, Matplotlib \citep{Hunter:2007}, Pandas \citep{Pandas:2010}.






\bibliography{references, refs-nonads}

\appendix

\end{document}